\documentclass[preprint]{aastex631}%631

\usepackage{graphicx}				
\usepackage{latexsym}
\usepackage[T1]{fontenc}
\usepackage{amsmath}
\usepackage{amsfonts}	
\usepackage{amssymb}
\usepackage{scalerel}
\usepackage{lmodern}
\usepackage{gensymb}
\usepackage{mdframed}
\usepackage{silence}

\begin{document}

\title{Using cGANs for Anomaly Detection: Identifying Astronomical Anomalies in JWST Imaging}

\author[0000-0002-0786-7307]{Ruby Pearce-Casey}
\affiliation{School of Physical Sciences \\
The Open University, Milton Keynes, MK7 6AA UK}

\author{Hugh Dickinson}
\affiliation{School of Physical Sciences \\
The Open University, Milton Keynes, MK7 6AA UK}

\author{Stephen Serjeant}
\affiliation{School of Physical Sciences \\
The Open University, Milton Keynes, MK7 6AA UK}

\author{Jane Bromley}
\affiliation{School of Computing \\
The Open University, Milton Keynes MK7 6AA UK}

\submitjournal{AAS Journal}

\section{Abstract} \label{sec:intro}

We present a proof of concept for mining JWST imaging data for anomalous galaxy populations using a \emph{conditional Generative Adversarial Network} (cGAN). We train our model to predict long wavelength NIRcam fluxes ($LW:F277W, F356W, F444W$ between $2.4\mu m - 5.0\mu m$) from short wavelength fluxes ($SW:F115W, F150W, F200W$ between $0.6\mu m - 2.3\mu m$) in $\sim2000$ galaxies. We test the cGAN on a population of 37 Extremely Red Objects (EROs) discovered by the CEERS JWST Team 
\citep{barro2023extremely}. Despite their red long wavelength colours, the EROs have blue short wavelength colours ($F150W-F200W \sim 0\,\text{mag}$) indicative of bimodal SEDs. Surprisingly, given their unusual SEDs, we find that the cGAN accurately predicts the \emph{LW} NIRcam fluxes of the EROs. However, it fails to predict \emph{LW} fluxes for other rare astronomical objects, such as a merger between two galaxies, suggesting that the cGAN can be used to detect some anomalies.\footnote{The source code and trained model are available at \url{https://github.com/RubyPC/Anomaly_Detection_with_cGANs}.}

\section{Method} \label{sec:method}
We use public domain JWST NIRcam data in the CEERS field, which covers $100sq.arcmin$ of the EGS field, and our network is based on the \emph{pix2pix} architecture described in \cite{8100115}. The data for the model were obtained from the CEERS data release $0.5$. Due to the different sensitivities of Module A and B of the JWST imaging, this study focussed on the subset of galaxies that were imaged using Module A. The images for each NIRcam filter were cropped into two regions of the NIRcam imaging field and a PSF-matching was done to match the PSF of each filter to that of F444W. Sources were detected using Python's \emph{sep} package\footnote{https://sep.readthedocs.io/en/v1.1.x/}\citep[]{Barbary2016,1996A&AS..117..393B} using $F277W$ as the detection filter. A pixel threshold value $\geq 2.0\sigma$ was used with $\sigma$, the global background \text{RMS}. This detection threshold is the minimum signal-to-noise ($S/N$) ratio of a pixel to be regarded as a detection. All detected sources were extracted as $90\times90$ pixel cutout images using Python's \emph{Astropy} package\footnote{https://readthedocs.org/projects/astropy/}\citep{astropy:2013, astropy:2018, astropy:2022}.

The cGAN was pre-trained on \emph{SW} NIRcam filters for a subset of sources detected to predict \emph{LW} filters using $90\%$ of the $\sim2000$ galaxies for training and $10\%$ for validation. After training the cGAN for $100$ epochs, the validation dataset was used to assess the generalisation capability of the model. To assess the performance of the cGAN, photometry was measured on the predicted \emph{LW} fluxes to compare the SED of the objects with those predicted by the model. Figure \ref{fig:plot}a shows the distribution of the absolute difference between the ground truth SED, $y$, and the SED predicted by the cGAN, $\hat{y}$, against a subset of the validation data set across the three \emph{LW} NIRcam filters. The absolute difference is predominantly situated at $\sim0$ for the majority of sources for all of the \emph{LW} filters. The distribution is roughly uniform for the $F356W$ filter where there is a small difference between the ground truth SED and the predicted SED, respectively. Nevertheless, this shows that the model can accurately predict NIRcam \emph{LW} filters such that we can reliably construct an SED.

\section{Results} \label{sec:results}
\begin{figure}[h!]
    \begin{mdframed}
    \centering
    \includegraphics[width=1.02\textwidth, scale=1.5]{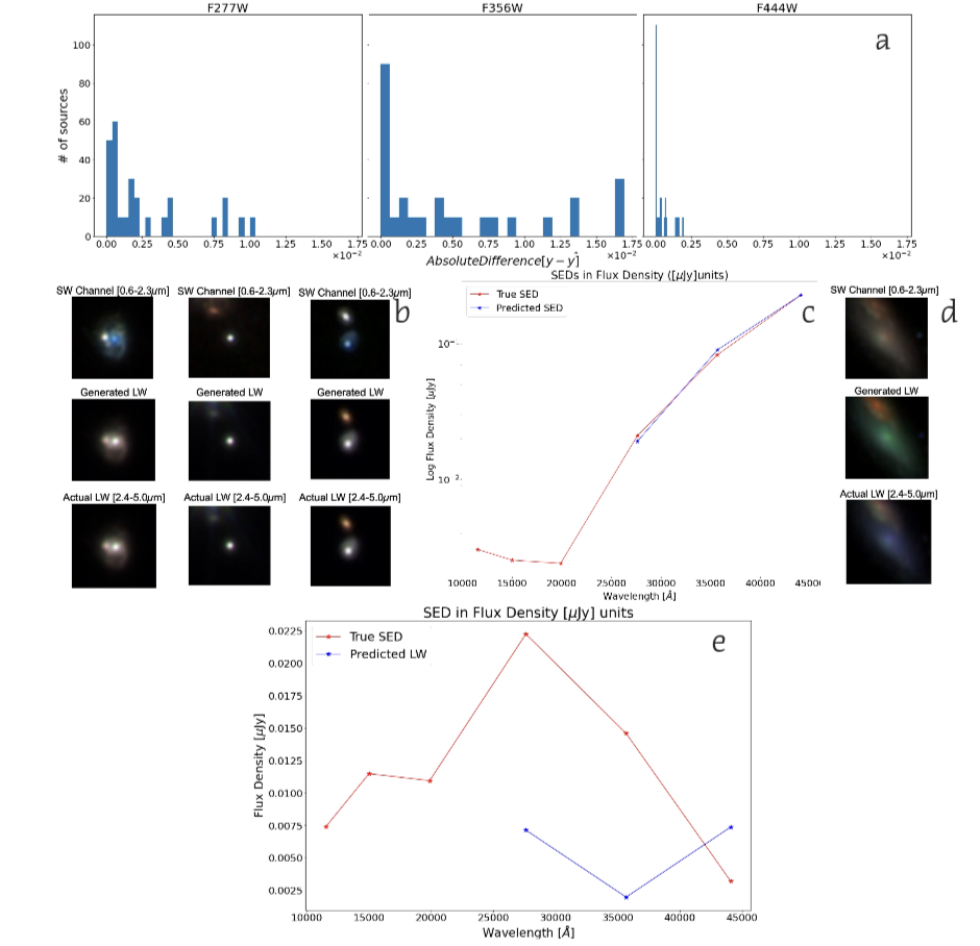}
    \end{mdframed}
    \caption{\emph{The results of the $cGAN$. \ref{fig:plot}\text{a} shows the distribution of the absolute difference in SED between the ground truth and the $cGAN\textquotesingle s$ prediction, respectively, after testing on a subset of the validation data set of $\sim2000$ objects. \ref{fig:plot}\text{b} shows the $cGAN\textquotesingle s$ prediction of the \emph{LW} channel for a set of $3$ EROs in the middle row comparing to the ground truth in the bottom row with the top row being the \emph{SW} input to the model, and \ref{fig:plot}\text{c} illustrates the $cGAN\textquotesingle s$ prediction for one ERO, showing its generalisation power of predicting the \emph{LW} fluxes of objects with peculiar SEDs by excellently constructing the characteristic steep rise in the $SED$. Finally, \ref{fig:plot}\text{d} shows the $cGAN\textquotesingle s$ poor prediction on a pair of merging galaxies with \ref{fig:plot}\text{e} showing the poorly constructed SED, where the cGAN can be used as an anomaly detector.}}
    \label{fig:plot}
\end{figure}

We tested the cGAN on the $37$ EROs identified by \cite{barro2023extremely} with $F444W < 28\,\text{mag}$ and photometric redshifts $5<z<7$. The $37$ EROs were deliberately excluded from the model\textquotesingle s training dataset. All of the EROs are unresolved, point-like sources in all observed filters and are typically faint in all \emph{SW} NIRcam filters. \cite{barro2023extremely} found that the EROs exhibit a distinctive, bimodal SED with blue colours for $\lambda > 2\mu m$ and they suggest that the flux in some of the \emph{LW} filters is partially boosted by strong emission lines, making the colours redder than the stellar continuum. A defining feature of these EROs is that \emph{all} of them have blue colours in the \emph{SW} NIRcam filters ($F115W-F200W \sim 0$). Figure \ref{fig:plot}b shows the cGAN\textquotesingle s prediction on $3$ of the EROs with the top row being the \emph{SW} input to the model and the middle and bottom rows being the cGAN\textquotesingle s prediction of the \emph{LW} and the ground truth \emph{LW}, respectively. Despite the peculiar, bimodal SEDs of the EROs, the cGAN performs just as well in predicting the \emph{LW} filters for a sample of EROs, predicting red \emph{LW} NIRcam fluxes. Additionally, Figure \ref{fig:plot}c shows the comparison between the ground truth SED of an identified ERO and the corresponding prediction from the cGAN. Although EROs are rare, and with a peculiar SED, the cGAN accurately predicts the steep rise in the SED at longer NIR wavelengths.

However, we find that the cGAN cannot accurately predict \emph{LW} filters for \emph{all} galaxy types. In particular, we find its predictions for galaxies with unusual or disturbed morphologies are less accurate, such as can be seen in Figure \ref{fig:plot}d where the prediction of the \emph{LW} in the middle row is not accurate in comparison to the ground truth in the bottom row for a pair of merging galaxies. Figure \ref{fig:plot}e shows the construction of the SED for the galaxy merger and it is clear that the model fails to predict the \emph{LW} NIRcam fluxes. 
Thus, we suggest that the cGAN can be used as an anomaly detector; it can accurately predict \emph{LW} filters for the majority of galaxies in the CEERS field, including EROs, but its performance suffers when predicting on rare objects such as a merger between galaxies. 

\newpage
\vspace{10pt}
\section{Conclusion}\label{sec: conclusion}
In conclusion, our results demonstrate our cGAN\textquotesingle s capability of detecting some rare astronomical objects. Therefore, for this specific case  using JWST NIRcam filters, not only can cGANs be used for image generation and image segmentation tasks but also for anomaly detection. This technology can be deployed on the rapidly increasing archive of JWST imaging survey observations as well as forthcoming giant survey data sets from e.g. Roman and Euclid; anomalous systems such as strong gravitational lenses have already been seen in JWST surveys \citep[e.g.][]{Pearson+23,vanDokkum+23}.

\section{Acknowledgements}
Ruby Pearce-Casey thanks the Science and Technology Facilities Council for support under grant ST/W006839/1.
This work made use of Astropy: a community-developed core Python package and an ecosystem of tools and resources for astronomy, and Sep\software{Astropy \citep{astropy:2013, astropy:2018, astropy:2022}}. \software{Sep \citep[]{Barbary2016,1996A&AS..117..393B}}.

\end{document}